\documentclass[trackchanges,twocolumn]{aastex701}

\begin{document}

\title{The Search for DB Double White Dwarf Binary Candidates from DESI, SDSS, and LAMOST}

\correspondingauthor{Jingkun Zhao, Huahui Yan}
\author{Keyan Li}
\altaffiliation{These authors contributed equally to this work.}
\affiliation{National Astronomical Observatories, Chinese Academy of Sciences, Beijing 100101,  People’s Republic of China}
\affiliation{Lanzhou University, Lanzhou 730000, People's Republic of China}

\email[]{}  

\author[orcid=0009-0005-8565-0858]{Huahui Yan}
\altaffiliation{These authors contributed equally to this work.}
\affiliation{School of Physics and Astronomy, Sun Yat-sen University, DaXue Road, Zhuhai 519082, GuangDong, People’s Republic of China.}
\affiliation{CSST Science Center for the Guangdong-Hongkong-Macau Greater Bay Area, Sun Yat-sen University, DaXue Road, Zhuhai 519082, GuangDong, People’s Republic of China}
\email[show]{yanhh8@mail2.sysu.edu.cn}  

\author{Jingkun Zhao}
\affiliation{National Astronomical Observatories, Chinese Academy of Sciences, Beijing 100101,  People’s Republic of China}
\email[show]{zjk@nao.cas.cn}

\author{Gang Zhao}
\affiliation{National Astronomical Observatories, Chinese Academy of Sciences, Beijing 100101,  People’s Republic of China}
\email[show]{}

\begin{abstract}
	Double white dwarfs (DWDs) are candidate progenitors of Type Ia
	supernovae and promising Galactic gravitational wave sources. Previous
	large-scale radial velocity (RV) variability searches for DWDs have been
	restricted almost exclusively to hydrogen-atmosphere (DA) white dwarfs
	(WDs), leaving the helium-atmosphere (DB) essentially unexplored.
	We present the first systematic search for RV-variable DB WDs and DB DWD
	binary candidates. Our initial sample is taken from the DB WD catalogue
	classified from DESI DR1, which we crossmatched with
	LAMOST DR13 and SDSS DR20 (including the SDSS-V daily spectra) to collect multi-epoch spectra. We measured RVs from the spectra
	by template fitting, and assessed the significance of RV variability with
	a $\chi^{2}$-based test, selecting candidates with $-\log p > 3$. From
	906 DB WDs with at least two valid RV measurements, we identified 9 DB
	binary candidates. None of them has been reported as a binary system
	before. Monte Carlo simulations show that our search is most sensitive to
	orbital periods of $\sim$0.5--10~h. Follow-up
	spectroscopic observations are required to confirm these candidates and
	to constrain their orbital parameters.
\end{abstract}



\section{Introduction}
\label{sec:intro}

White dwarfs (WDs) are the end products of the evolution of stars with
initial masses below $\sim$8--10 $M_{\odot}$ \citep{ Doherty2014}.
Since their progenitors span nearly the entire mass range of star formation,
more than 95\% of all stars in the Galaxy will eventually become WDs
\citep{Fontaine2001,Lauffer2018}. Their interiors cool along
slowly evolving, well-understood sequences, so WDs are widely used as
chronometers and tracers of the star formation history of the Milky Way
\citep{Torres2005, Tremblay2014}. Spectroscopy, photometry, and
asteroseismology give access to their key physical properties, such as mass,
cooling age, and the composition of their atmospheres and interiors
\citep{Koester2009,Bergeron2011}.

Based on the dominant features of their optical spectra, WDs are classified
into several spectral types \citep{Sion1983,Koester2010}.
About 80\% of them show only hydrogen absorption lines and are grouped
into the DA class. Most of the remaining WDs that are hot enough to excite
neutral helium display spectra dominated by \ion{He}{1} lines and are
classified as DB WDs, the most common non-DA type, while cooler helium-rich
objects showing metals or carbon are labeled DZ and DQ, respectively.

Type Ia supernovae (SNe~Ia) serve as standardizable candles in cosmology,
and their use led to the discovery of the cosmic acceleration
\citep{Riess1998, Perlmutter1999}. These events are generally attributed to
the thermonuclear disruption of a WD in a binary system, with two main
progenitor scenarios: the single-degenerate scenario, in which a WD accretes
matter from a non-degenerate companion \citep{Whelan1973, Nomoto1984}, and
the double-degenerate scenario, in which the merger of two WDs in a close
binary triggers the explosion \citep{Webbink1984, Iben1984, Maoz2014,
	Liu2023}. Beyond being candidate SN~Ia progenitors, close double white
dwarfs (DWDs) are among the most promising Galactic gravitational-wave
sources for the Laser Interferometer Space Antenna (LISA)
\citep{AmaroSeoane2023}, and they provide key
constraints on common-envelope evolution and binary interaction physics
\citep{Han1995}. However, only a few hundred DWDs have been
confirmed so far, far below the expected several per cent of the WD
population \citep{Maxted2000, Maxted2002, Rebassa2010,Napiwotzki2020, Munday2024}, because WDs are
faint and their broad absorption lines demand multi-epoch spectroscopy to
reveal orbital motion.

Measuring radial velocity (RV) variability has long been one of the most
efficient ways to uncover DWDs. Early searches were carried out on small
samples \citep{Saffer1988, Bragaglia1990, Marsh1995, Maxted2000}, and the dedicated ESO
SN~Ia Progenitor Survey \citep[SPY;][]{Napiwotzki2020} systematically
monitored about one thousand WDs, while the Extremely Low Mass WD survey
targeted He-core WDs \citep{Brown2010}. In the era of large
spectroscopic surveys, RV-variability searches have been extended to much
larger samples: \citet{Yan2024} identified 65 DA DWD candidates from SDSS
DR14, \citet{Jiang2025} found 33 DA DWD candidates based on the DESI Early
Data Release, and \citet{Adamane2026} presented 63 DA DWD candidates
from SDSS-V DR19 and constrained the close-binary fraction of WDs. We note,
however, that all of these large-scale searches were restricted to DA WDs.
Although the number of spectroscopically confirmed DB WDs has grown to about
one thousand thanks to SDSS \citep{Kepler2019}, no
systematic RV-variability search has ever been performed for DB WDs, even
though their \ion{He}{1} lines are intrinsically narrower than the
pressure-broadened Balmer lines of DAs and therefore allow precise RV
measurements. The DB  thus remains essentially unexplored in the
search for DWDs.

This situation can now be improved with the latest data releases of large
spectroscopic surveys. SDSS-V (DR20) is obtaining multi-epoch spectroscopy
for the first large WD sample of the \textit{Gaia} era, with several
sub-exposures per target \citep{Kollmeier2017, Adamane2026}. The Large Sky Area Multi-Object Fiber Spectroscopy Telescope (LAMOST)
survey has released more than ten million low-resolution stellar spectra \citep{Cui2012, Zhao2012}, and the Dark Energy Spectroscopic
Instrument (DESI) has obtained spectra of tens of millions of targets in its
second data release (DR1), including a large number of WDs
\citep{DESI2016, DESI2022}. Crossmatched with the
\textit{Gaia} DR3 WD catalog \citep{Gentile2021}, these data sets
provide an unprecedented opportunity to search for DB WDs and to investigate
their RV variability over a large fraction of the sky.

In this paper, we present the first systematic search for RV-variable DB
WDs. Our initial sample is taken from the catalog of DB WDs classified from
DESI DR1 spectra by \citet{swan2026desidatarelease1}. We crossmatch this sample with the
spectroscopic data of LAMOST DR13 and SDSS DR20 to collect multi-epoch
spectra for each target. 

This paper is organized as follows. In Section~\ref{sec:data}, we describe
the survey data. Section~\ref{sec:method} presents
the identification of DB candidates and the measurements of RVs and
atmospheric parameters. In Section~\ref{sec:selection}, we present the DB
sample and the RV-variable binary candidates. We discuss our results 
and summarize in Section~\ref{sec:conclusion}.

\section{Data}
\label{sec:data}

\subsection{Initial DB sample}
\label{sec:sample}

Our initial sample is drawn from the white dwarf (WD) catalogue of
\citet{swan2026desidatarelease1}, which was constructed from the second data release of the
DESI. In that
work, the spectra of 63,968 WD candidates were visually inspected and
fitted with WD atmosphere models, yielding 44,409 spectroscopically
confirmed WDs. Rather than assigning a single definitive label, each
spectrum was classified independently by at least two inspectors, and
disagreements (about 10 per cent of the objects after the first round) were
resolved through additional rounds of inspection and verification, with a
control sample of randomly exchanged spectral types used to estimate the
random error rate. Confidence scores were assigned to the resulting
classifications, allowing statistically clean samples to be selected. We
selected all objects whose spectral type is classified as DB as our
initial DB WD sample.

\subsection{Multi-epoch spectroscopy from LAMOST and SDSS}
\label{sec:lamost_sdss}

To obtain multi-epoch spectroscopy for the initial sample, we crossmatched
it with  LAMOST DR13 and SDSS DR20 adopting a
matching radius of 3 arcsec. For matched targets, we downloaded all
available spectra. LAMOST obtains low-resolution spectra
($R \sim 1800$) covering the optical range, and repeated observations of
the same target naturally provide multiple epochs. For SDSS, we downloaded
all spectra obtained before SDSS-V, together with the SDSS-V DR20 daily
spectra. Combining these two products maximizes the temporal baseline for
the RV variability analysis presented in Section~\ref{sec:method}. 

\subsection{DESI  quality cuts}
\label{sec:desi_rv}

For the DESI data, we adopted the per-exposure RV measurements of
\citet{swan2026desidatarelease1}, who measured RVs independently for the blue (B) and red
(R) spectrograph arms by maximum-likelihood fitting of 32-\AA\ windows
centred on 23 \ion{He}{1} lines, with a Gaussian prior of width
1000~$\mathrm{km\,s^{-1}}$ centred on 30~$\mathrm{km\,s^{-1}}$ and no
correction for gravitational redshift. Since some per-exposure measurements
are unreliable, we selected the reliable ones following \citet{swan2026desidatarelease1},
requiring each exposure to satisfy three criteria: (i) RV uncertainties
below 50~$\mathrm{km\,s^{-1}}$ in both arms (relaxed from their
30~$\mathrm{km\,s^{-1}}$ limit to retain more single-exposure data; failed
fits are returned with $\sim$1000~$\mathrm{km\,s^{-1}}$ uncertainties and
are thus excluded automatically); (ii) $|RV| < 750~\mathrm{km\,s^{-1}}$ in
both arms, the threshold they used to separate genuine velocities from
failed fits; and (iii) agreement between the two arms within three times
their combined uncertainty,
\begin{equation}
	|RV_{B} - RV_{R}| < 3\sqrt{\sigma_{B}^{2} + \sigma_{R}^{2}},
	\label{eq:arm_consistency}
\end{equation}
which removes a further $\sim$1.5 per cent (180) of the measurements,
mostly affected by wavelength-calibration or masking problems in one arm.

For each valid exposure, we combined the two arms with the
inverse-variance-weighted average adopted by \citet{swan2026desidatarelease1}:
\begin{equation}
	RV = \frac{RV_{B}/\sigma_{B}^{2} + RV_{R}/\sigma_{R}^{2}}
	{1/\sigma_{B}^{2} + 1/\sigma_{R}^{2}}, \qquad
	\sigma_{RV} = \left( \frac{1}{\sigma_{B}^{2}} +
	\frac{1}{\sigma_{R}^{2}} \right)^{-1/2}.
	\label{eq:rv_combine}
\end{equation}
We adopted this combined value as the RV of each DESI multi-epoch
observation.

\section{Radial velocity measurements}
\label{sec:method}

We measured RVs from the LAMOST and SDSS spectra with a modified version
of the \texttt{corv}\footnote{\url{https://github.com/vedantchandra/corv}}
package \citep{Arseneau2024}. Both surveys provide vacuum wavelengths
already reduced to the heliocentric frame, so no additional wavelength
correction was applied. For each spectrum, we retained only pixels with a
positive inverse variance and a zero bad-pixel mask, and adopted
$\sigma_{f} = \mathrm{ivar}^{-1/2}$ as the flux uncertainty.

The observed line profiles were fitted with the helium-dominated
three-dimensional model atmosphere grid of \citet{Cukanovaite2021}, using the \ion{He}{1} lines at 4471, 4921,
5875, and 6678\,\AA. Each line was continuum-normalized independently
before fitting: we extracted a window of half-width
$\Delta\lambda = 50$--80\,\AA\ around the rest wavelength, took a fixed
number of pixels (20--40, depending on the line) at each end of the window
as continuum, and fitted a first-order polynomial to these two edge
regions.

\begin{figure*}[ht!]
	\centering
	\includegraphics[width=\textwidth]{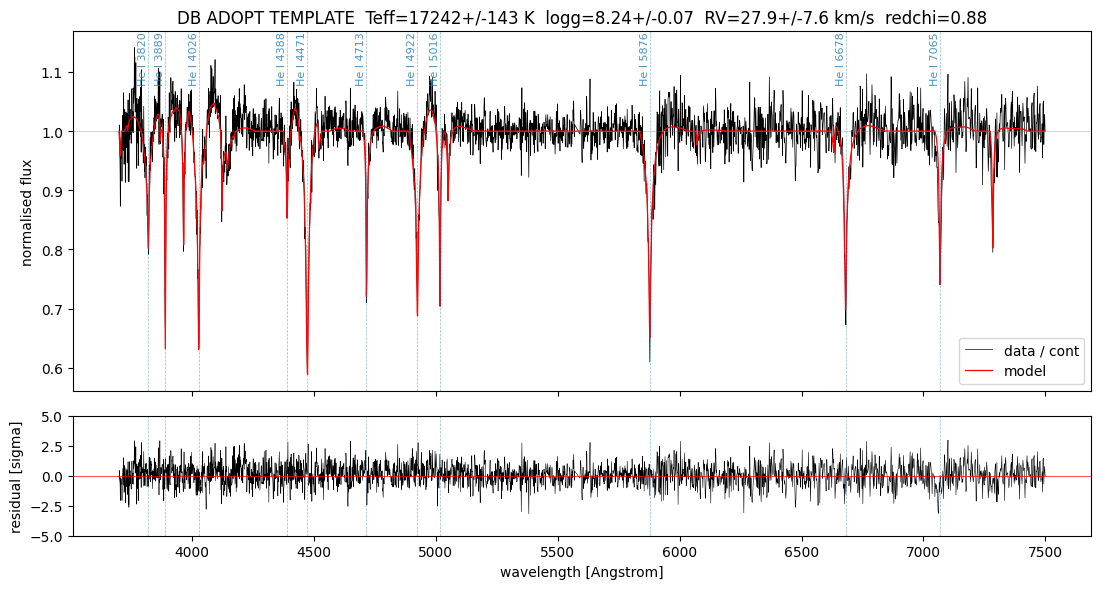}
	\caption{Example of the template fit for the DB WD
		WDJ234459.42+024109.11. Top panel: the normalized observed spectrum
		(black) and the normalized best-fitting helium-atmosphere model (red).
		The dashed vertical lines mark the \ion{He}{1} lines used in the fit,
		and the best-fitting parameters ($T_{\mathrm{eff}} = 17242 \pm
		143$~K, $\log g = 8.24 \pm 0.07$, $RV = 27.9 \pm
		7.6~\mathrm{km\,s^{-1}}$, with a reduced $\chi^{2}$ of 0.88) are given
		at the top. Bottom panel: the fit residuals in units of the flux
		uncertainty ($\sigma$). }
	\label{fig0}
\end{figure*}

The normalized profiles of all lines were then fitted simultaneously with
\texttt{lmfit}, with the effective temperature, surface gravity, and RV as
free parameters and no external priors on the atmospheric parameters.
Holding the best-fitting template fixed, the velocity was refined by
evaluating $\chi^{2}$ on a grid of 500 velocities spanning
$\pm1500~\mathrm{km\,s^{-1}}$ and fitting a parabola to the region within
300~$\mathrm{km\,s^{-1}}$ of the minimum. The vertex of the parabola gives
the final velocity, and the half-width of the $\Delta\chi^{2} = 1$
interval gives its uncertainty. Since the template is held fixed in this step, the RV is the only free
parameter, and $\Delta\chi^{2} = 1$ corresponds to the $1\sigma$
confidence interval. These template-fitting velocities are
adopted throughout this work.  Table \ref{tab1} shows the  RVs and atmospheric parameters of the DB WD sample derived from template fitting. An example of the fit is shown in Figure~\ref{fig0} for the DB WD
WDJ234459.42+024109.11, where the black line is the normalized observed
spectrum and the red line is the normalized best-fitting model.

\begin{figure*}[ht!]
	\centering
	\includegraphics[width=\textwidth]{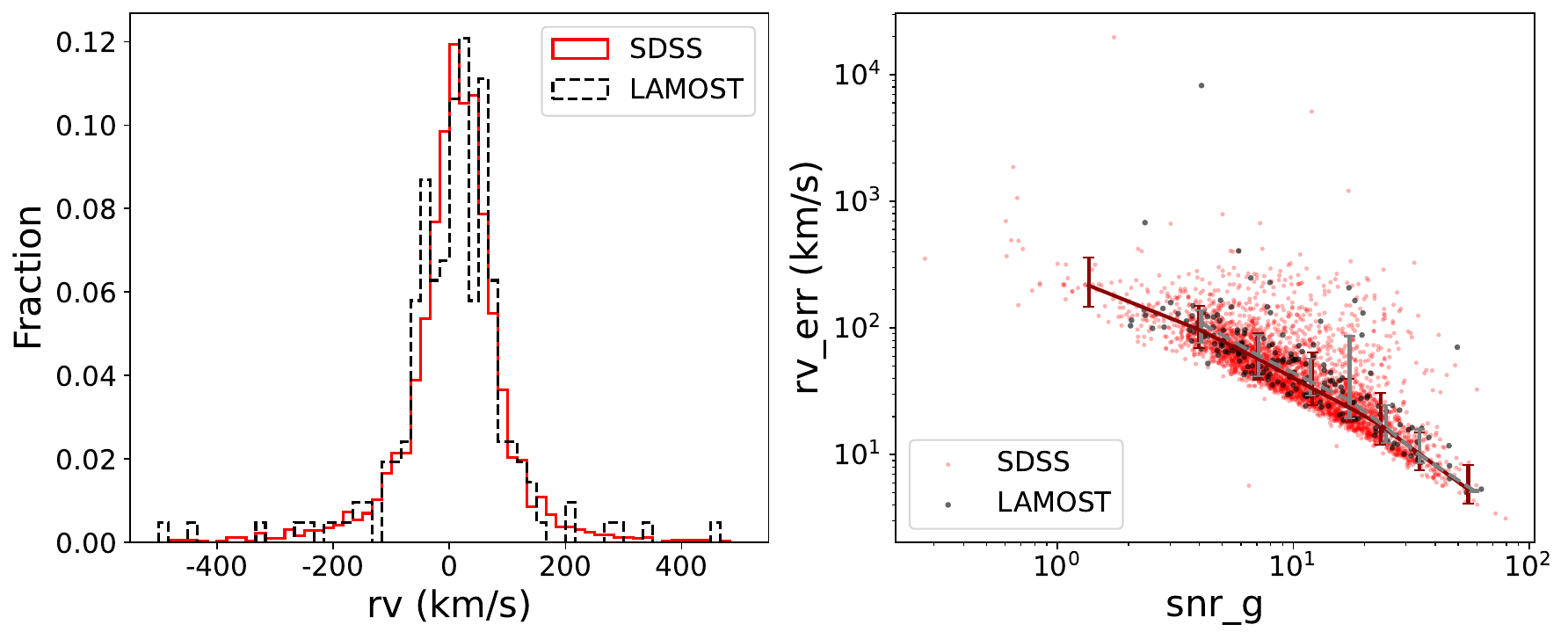}
	\caption{Left panel: RV distributions of the final SDSS and
		LAMOST  samples. Both distributions peak near zero and are
		concentrated within $\pm100~\mathrm{km\,s^{-1}}$, with only a small
		tail of larger $|RV|$ measurements. Right panel: RV uncertainty as a
		function of the $g$-band signal-to-noise ratio (S/N$_{g}$) for the two
		samples. The uncertainty decreases with S/N$_{g}$ as an approximate
		power law with a slope close to $-1$, indicating that the measurement
		precision is dominated by photon noise.}
	\label{fig1}
\end{figure*}

Figure~\ref{fig1} shows the RV distributions and the RV uncertainties as a
function of the $g$-band signal-to-noise ratio (S/N$_{g}$) for the final
LAMOST and SDSS spectra. The RV distributions of the two samples are
similar: both are concentrated within $\pm100~\mathrm{km\,s^{-1}}$, peak
near zero, and show only a small tail of larger $|RV|$ measurements. The
RV uncertainty decreases with S/N$_{g}$ as an approximate power law with a
slope close to $-1$,
indicating that the precision is dominated by photon noise: the median
uncertainty is $\sim$10~$\mathrm{km\,s^{-1}}$ at S/N$_{g} > 30$, but
rises above $\sim$80~$\mathrm{km\,s^{-1}}$ at S/N$_{g} < 5$. We discarded
spectra for which the fit did not converge, and required $\sigma_{RV} <
100~\mathrm{km\,s^{-1}}$ and $|RV| < 300~\mathrm{km\,s^{-1}}$ to exclude
extreme measurements. We verified that this cut does not remove genuine candidates: relaxing or
removing it restores only 23 spectra of 19 targets, all 9 candidates presented
in Section~\ref{sec:rvvar} are retained, and visual inspection shows that the
additional formal detections are caused by low-S/N spectra or problematic fits
rather than coherent velocity shifts. The final sample contains 2667 SDSS spectra and 169
LAMOST spectra.

\section{Selection of RV-variable targets}
\label{sec:selection}

\subsection{RV error test}
Before selecting RV-variable targets, we first verified the reliability of
our RV uncertainties, following the approach of \citet{Maoz2017}. The idea
is straightforward: most WDs in the sample are single stars with constant
RVs, so for any two exposures of the same target, the RV difference
$\Delta RV$ should arise purely from measurement noise. Since the noise of
the two measurements is independent, $\Delta RV$ is expected to follow a
Gaussian distribution centred on zero, with a standard deviation equal to
the combined error $\delta RV = \sqrt{\sigma_{i}^{2} + \sigma_{j}^{2}}$.
Only the small fraction of genuine binaries is expected to deviate from
this behaviour, producing extended tails in the distribution. We therefore
grouped all pairs of exposures of the same target into five bins of
$\delta RV$ (20~$\mathrm{km\,s^{-1}}$ wide, covering
0--100~$\mathrm{km\,s^{-1}}$), and fitted the $\Delta RV$ distribution in
each bin with a Gaussian. To prevent the binary tails from biasing the
fits, the fitting was initialised with robust estimates based on the
median absolute deviation. Comparing the measured scatter
$\sigma_{\mathrm{fit}}$ with $\delta RV$ then tests whether the formal
errors are underestimated or overestimated.

\begin{figure*}[ht!]
	\centering
	\includegraphics[width=\textwidth]{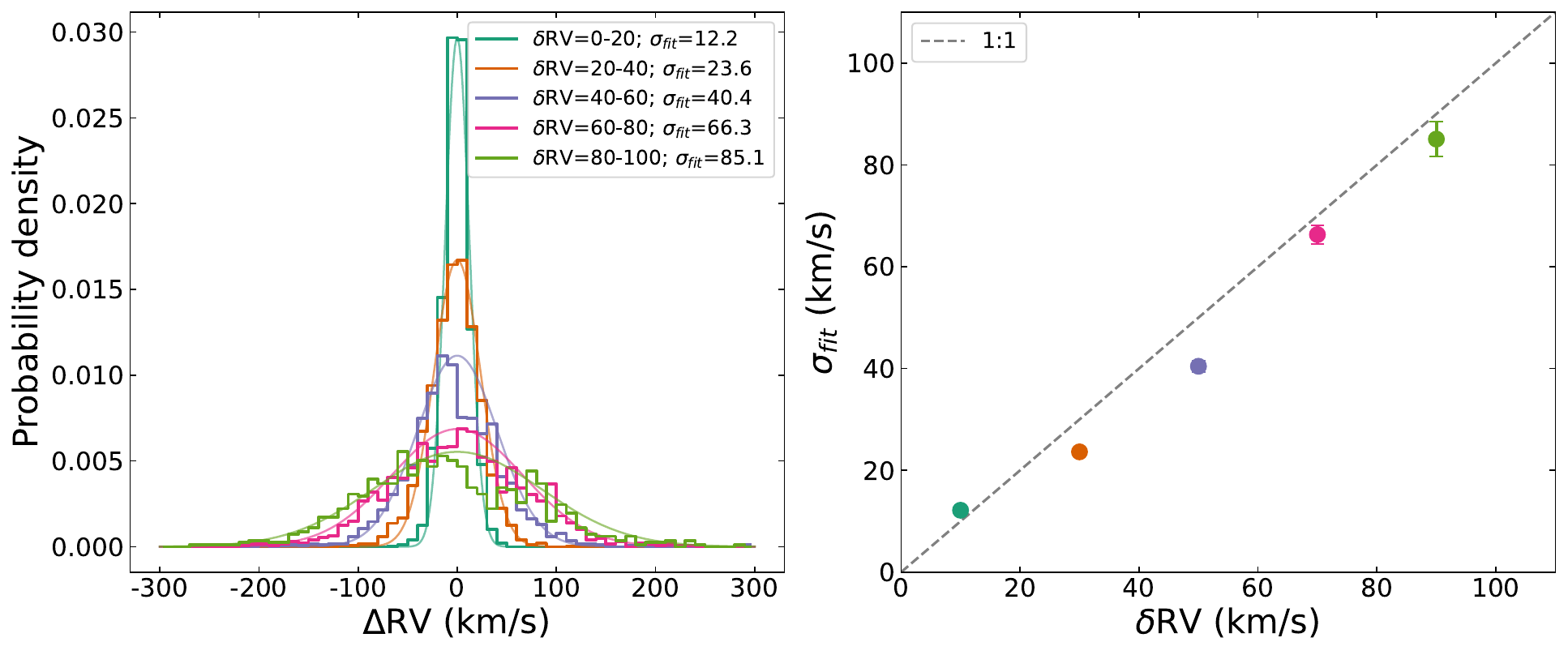}
	\caption{Validation of the RV uncertainties using pairs of exposures
		of the same target. Left panel: distributions of the RV difference $\Delta
		RV$ for exposure pairs, grouped into five bins of the combined error
		$\delta RV = \sqrt{\sigma_{i}^{2}+\sigma_{j}^{2}}$
		(20~$\mathrm{km\,s^{-1}}$ wide). Each normalised distribution (step
		lines) is shown with its Gaussian fit (smooth curves), and the fitted
		scatter $\sigma_{\mathrm{fit}}$ is given in the legend. Right panel:
		comparison of $\sigma_{\mathrm{fit}}$ with the central $\delta RV$ of
		each bin; the grey dashed line marks the one-to-one relation. The
		$\sigma_{\mathrm{fit}}$ values of all bins lie close to the one-to-one
		line, indicating that the formal RV uncertainties are reliable.}
	\label{fig2}
\end{figure*}

The result is shown in Figure~\ref{fig2}. The $\sigma_{\mathrm{fit}}$
values of all five bins lie close to the one-to-one relation, with no
systematic offset, indicating that our RV uncertainties are accurate and
that no significant additional noise source is present. We note that the
pairs include measurements from different instruments (DESI, SDSS, and
LAMOST); the good agreement therefore also suggests that any RV zero-point
offsets between the three surveys are small compared to the measurement
uncertainties.

As a further check, we computed the median RV difference of
same-target pairs separately for each combination of instruments.
The resulting zero-point offsets are all within
$\pm$10~$\mathrm{km\,s^{-1}}$ ($-7.0$~$\mathrm{km\,s^{-1}}$
for DESI--SDSS and $+0.3$~$\mathrm{km\,s^{-1}}$ for LAMOST--SDSS
pairs), far smaller than the RV differences of
50--440~$\mathrm{km\,s^{-1}}$ shown by our candidates.
Cross-instrument zero-point offsets therefore cannot explain any of
the selected candidates, including those whose epochs come from
different instruments.

\subsection{Selection of RV-variable targets}
\label{sec:rvvar}

We assessed the significance of RV variability for each target using the
$\chi^{2}$-based method commonly adopted in previous DWD searches
\citep{Maxted2000, Napiwotzki2020, Yan2024, Adamane2026}. For each star, we first
computed the inverse-variance-weighted mean RV from all available epochs.
Assuming that the true RV is constant and equal to this mean, we then
computed the $\chi^{2}$ statistic, with $N-1$ degrees of freedom for $N$
epochs. Comparing the measured $\chi^{2}$ with the $\chi^{2}$
distribution for the corresponding degrees of freedom yields the
probability $p$ that random fluctuations around a constant velocity could
produce a $\chi^{2}$ equal to or larger than the observed value. We
considered a target to be significantly RV-variable when $-\log p > 3$.\footnote{We adopted this relatively low threshold to obtain a more complete list of
	candidates for follow-up observations, whereas a more conservative threshold of
	$-\log p > 4$ was used by \citet{Maxted2000} to identify RV-variable WDs.}

\begin{figure*}[ht!]
	\centering
	\includegraphics[width=\textwidth]{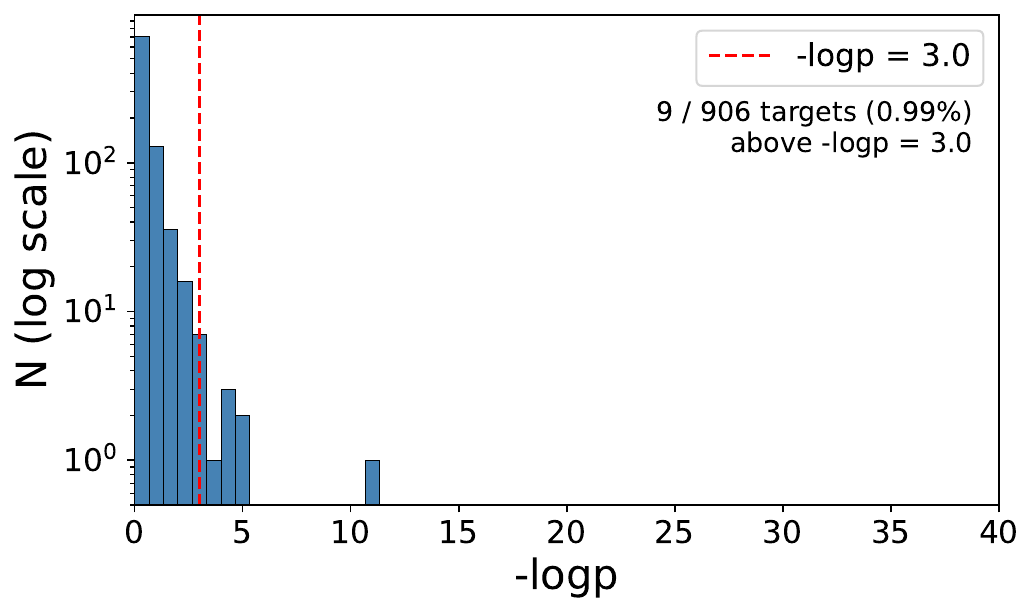}
	\caption{Histogram of $-\log p$ for the 906 DB WDs in our sample, where $p$
		is the probability that the observed RV variations arise from random
		fluctuations around a constant velocity. The dashed line indicates
		our selection threshold of $-\log p = 3$, and the 9 targets exceeding
		this threshold are selected as DB binary candidates.}
	\label{fig3}
\end{figure*}

\begin{figure*}[ht!]
	\centering
	\includegraphics[width=\textwidth]{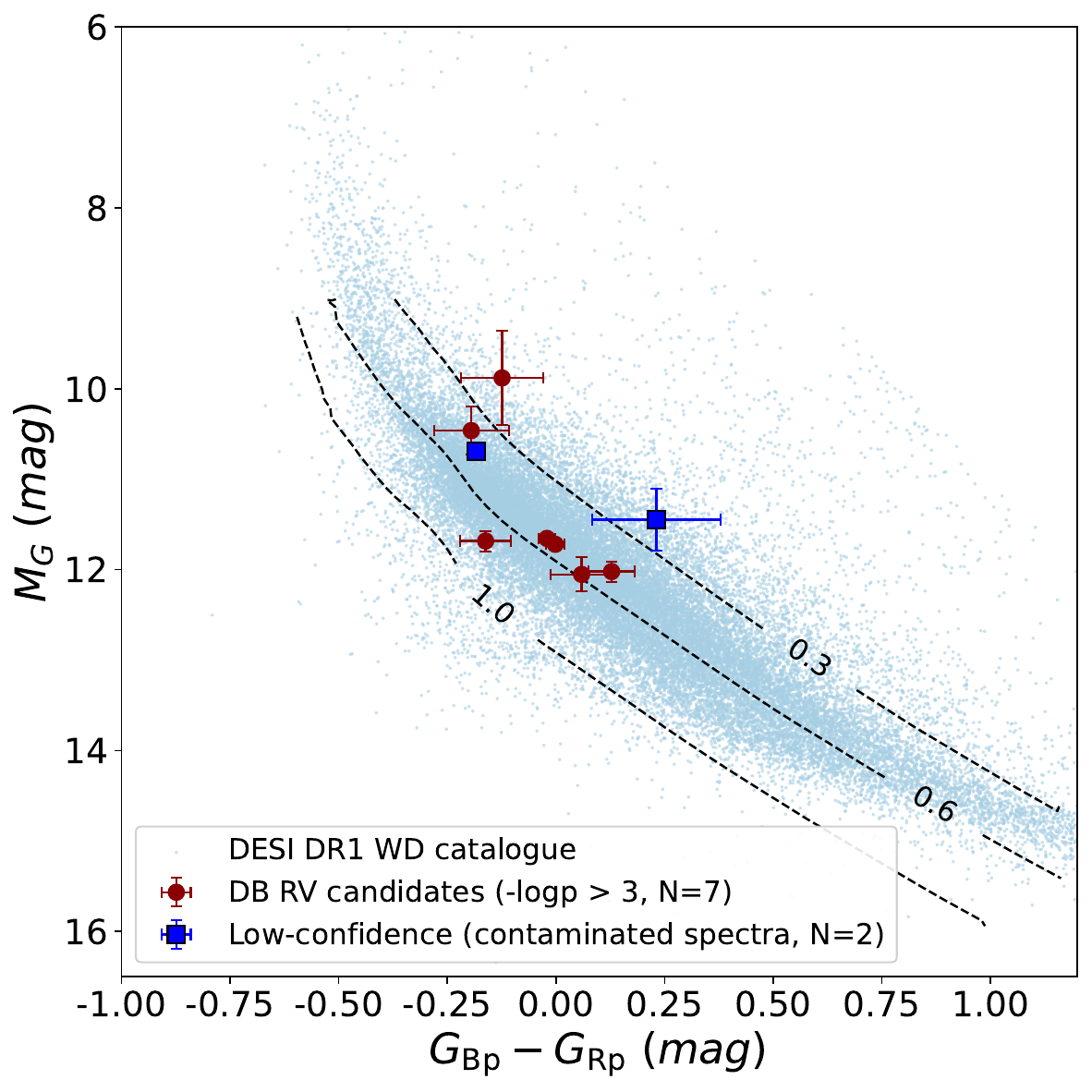}
	\caption{Gaia colour--magnitude diagram ($M_{G}$ versus
		$G_{\rm BP}-G_{\rm RP}$) of the DB binary candidates. The blue
		background points show all WDs in the DESI DR1 catalogue
		\citep{swan2026desidatarelease1}, the red circles mark the 7 RV-variable candidates identified in this work, and
		the blue squares mark the two low-confidence candidates
		(WDJ225336.79+081608.54 and WDJ085611.94+365850.63) whose spectra are
		contaminated by a wide M-dwarf companion. The black lines are WD cooling tracks
		of constant mass (0.3, 0.6, and 1.0~$M_{\odot}$). The candidates lie within the main locus of the
		WD population, with most of them around the typical WD mass of
		$\sim$0.6~$M_{\odot}$.}
	\label{fig4}
\end{figure*}

Applying this criterion to the 906 DB WDs in our sample, we identified 9
targets with significant RV variability, which we regard as DB binary
candidates. The distribution of $\log p$ for the full sample is shown in
Figure~\ref{fig3}, where these 9 candidates are clearly separated from the
bulk of the sample.  Their detailed information is listed in Table~\ref{tab2}. The positions of
the 9 candidates in the \textit{Gaia} colour--magnitude diagram are shown
in Figure~\ref{fig4}, where they are compared with the full DESI DR1 WD
sample and with WD cooling tracks of constant mass.

To check whether our candidates have been previously studied, we queried
the literature through SIMBAD \citep{Wenger2000} using
\texttt{astroquery} \citep{Ginsburg2019}. None of the 9 candidates has been
reported as a binary or binary candidate before. All of them are
therefore newly identified DB binary candidates. We also searched the Zwicky
Transient Facility \citep[ZTF;][]{Bellm2019, Graham2019, Masci2019}  and the Transiting Exoplanet Survey Satellite \citep[TESS;][]{Ricker2015} archival photometry of the 9 candidates
for light-curve variability. ZTF data are available for all of them and TESS
data for eight, but no secure photometric period is found.

Three candidates (WDJ225336.79+081608.54, WDJ085611.94+365850.63, and
	WDJ011532.11-004845.70) have a resolved Gaia DR3 neighbour within
	5~arcsec, at separations of 2.80, 2.74, and 4.72~arcsec, respectively. These
	neighbours are red, with $G_{\rm BP}-G_{\rm RP}$ between 1.9 and 2.7, and their
	parallaxes and proper motions are consistent with those of the WDs. The
	parallax differences are within $1.8\sigma$, and the proper-motion
	differences are below 1.35~mas~yr$^{-1}$, corresponding to tangential
	velocity differences of $\lesssim1.2~\mathrm{km\,s^{-1}}$. This suggests that
	the neighbours are likely wide, common proper motion K/M dwarf companions.
	Such wide orbits cannot produce measurable RV variations, but light from the
	companion can enter the spectroscopic fibre. We therefore re-examined the
	spectra of these candidates. For WDJ225336.79+081608.54, the two DESI spectra
	show no red excess and give $RV=-19~\mathrm{km\,s^{-1}}$, whereas the LAMOST
	spectrum taken on MJD 57327 is dominated by the M dwarf. This
	spectrum yields $RV=+276\pm40~\mathrm{km\,s^{-1}}$. Without this epoch, the
	remaining RVs are consistent with a constant velocity ($-\log p=0.23$). For
	WDJ085611.94+365850.63, the LAMOST spectrum is dominated by the M dwarf
	companion and gives $RV=+293\pm68~\mathrm{km\,s^{-1}}$, and the SDSS spectrum
	($RV=-12\pm22~\mathrm{km\,s^{-1}}$) also shows contamination at red
	wavelengths. The apparent RV variability of these two targets is therefore
	most likely caused by contamination from the companion, and we regard them as
	low-confidence candidates. WDJ011532.11$-$004845.70 also has a wide common
	proper motion companion, but its two SDSS-V spectra show no red excess, and
	the wide orbit cannot produce the observed RV difference. Its RV variability
	therefore cannot be attributed to the companion. If confirmed, it would
	indicate an unseen close companion to the WD in a hierarchical triple system.
	No red excess is found in the spectra of the other candidates.

\begin{deluxetable*}{lcccccrrcccll}
	\label{tab1}
	\tabletypesize{\small}
	\tablecaption{Radial velocities and atmospheric parameters of the DB white dwarf sample derived from template fitting.\label{tab:rv_template}}
	\tablewidth{0pt}
	\tablehead{
		\colhead{Target} & \colhead{RA} & \colhead{Dec} & \colhead{MJD} & \colhead{$v_r$} & \colhead{$\sigma_{v_r}$} & \colhead{$\chi^2_{\rm red}$} & \colhead{$T_{\rm eff}$} & \colhead{$\sigma_{T_{\rm eff}}$} & \colhead{$\log g$} & \colhead{$\sigma_{\log g}$} & \colhead{Spec} & \colhead{SDSS} \\
		\colhead{} & \colhead{(deg)} & \colhead{(deg)} & \colhead{} & \colhead{(km s$^{-1}$)} & \colhead{(km s$^{-1}$)} & \colhead{} & \colhead{(K)} & \colhead{(K)} & \colhead{} & \colhead{} & \colhead{} & \colhead{}
	}
	\startdata
	WDJ000255.14+124200.01 & 0.72985 & 12.70012 & 60563 & 66.73 & 46.46 & 0.699 & 19572 & 1619 & 7.972 & 0.399 & SDSS & dr20\_daily \\
	WDJ000255.14+124200.01 & 0.72986 & 12.70010 & 60564 & 57.83 & 57.28 & 0.674 & 23241 & 2517 & 8.383 & 0.321 & SDSS & dr20\_daily \\
	WDJ000515.55+071313.93 & 1.31493 & 7.22048 & 55860 & 34.76 & 10.97 & 0.905 & 18721 & 208 & 8.325 & 0.075 & SDSS & dr17 \\
	WDJ000550.92+041128.27 & 1.46211 & 4.19103 & 59905 & -54.49 & 87.98 & 0.761 & 15043 & 782 & 9.000 & 0.087 & SDSS & dr20\_daily \\
	WDJ000550.92+041128.27 & 1.46211 & 4.19104 & 60265 & -98.13 & 85.18 & 0.724 & 14799 & 907 & 7.070 & 1.328 & SDSS & dr20\_daily \\
	\multicolumn{13}{c}{$\cdots$} \\
	WDJ235443.44+014154.03 & 358.68112 & 1.69811 & 59485 & -8.27 & 9.85 & \nodata & \nodata & \nodata & \nodata & \nodata & DESI & \nodata \\
	WDJ235443.44+014154.03 & 358.68112 & 1.69811 & 59571 & -6.69 & 20.60 & \nodata & \nodata & \nodata & \nodata & \nodata & DESI & \nodata \\
	WDJ235703.87+061915.46 & 359.26620 & 6.32088 & 59501 & -21.53 & 13.67 & \nodata & \nodata & \nodata & \nodata & \nodata & DESI & \nodata \\
	WDJ235703.87+061915.46 & 359.26620 & 6.32088 & 59501 & -22.21 & 14.83 & \nodata & \nodata & \nodata & \nodata & \nodata & DESI & \nodata \\
	WDJ235703.87+061915.46 & 359.26620 & 6.32088 & 59529 & -15.00 & 12.08 & \nodata & \nodata & \nodata & \nodata & \nodata & DESI & \nodata
	\enddata
	\tablecomments{The full table is available in machine-readable form. For LAMOST and SDSS spectra, the RVs and the
		atmospheric parameters ($T_{\mathrm{eff}}$ and $\log g$) are derived from
		our template fitting; for DESI spectra, the
		RVs are the per-exposure measurements of \citet{swan2026desidatarelease1} after our
		quality cuts, for which no atmospheric
		parameters are given here.}
\end{deluxetable*}

\begin{deluxetable}{lrrrrlr}
	\label{tab2}
	\tabletypesize{\small}
	\tablecaption{RV variable candidates among the DB white dwarf sample with Gaia DR3 counterparts. $N$ is the total number of spectra, split by survey (LAMOST, SDSS, DESI), and $- \log \mathcal{P}$ is the logarithm of the false-alarm probability from the  $\chi^{2}$ test.}
	\tablewidth{0pt}
	\tablehead{
		\colhead{Target} & \colhead{$N$} & \colhead{$N_{\rm LAMOST}$} & \colhead{$N_{\rm SDSS}$} & \colhead{$N_{\rm DESI}$} & \colhead{$-\log \mathcal{P}$} & \colhead{Gaia DR3 source\_id} \\
		\colhead{} & \colhead{} & \colhead{} & \colhead{} & \colhead{} & \colhead{} & \colhead{}
	}
	\startdata
	WDJ225336.79+081608.54$^{\dagger}$ & 4 & 2 & 0 & 2 & 10.937 & 2713073189537921536 \\
	WDJ161643.66+171453.48 & 5 & 0 & 3 & 2 & 5.152 & 1198777431614139008 \\
	WDJ145023.39-005440.96 & 7 & 0 & 5 & 2 & 4.968 & 3650436912692498560 \\
	WDJ085611.94+365850.63$^{\dagger}$ & 2 & 1 & 1 & 0 & 4.655 & 718679076664002304 \\
	WDJ151907.02+464521.42 & 2 & 0 & 2 & 0 & 4.540 & 1588084533021284864 \\
	WDJ041912.68-034008.98 & 2 & 0 & 2 & 0 & 4.111 & 3203978257729004416 \\
	WDJ011532.11-004845.70 & 2 & 0 & 2 & 0 & 3.489 & 2533686798020139648 \\
	WDJ223255.41-084044.23 & 4 & 0 & 1 & 3 & 3.274 & 2621250392226389248 \\
	WDJ174810.29+452356.07 & 2 & 0 & 2 & 0 & 3.237 & 1350540234161336576
	\enddata
	\tablecomments{Candidates are selected with -$\log p > 3$ among sources with at least two spectra.$^{\dagger}$ Low-confidence candidates whose spectra are contaminated by a
		wide M-dwarf companion (see Section~\ref{sec:rvvar}).}
\end{deluxetable}

\subsection{Sensitivity to orbital period}
\label{sec:sensitivity}

To quantify the range of orbital periods to which our search is sensitive, we
performed Monte Carlo simulations of binaries observed with the same epochs and RV uncertainties as our sample, following \citet{Breedt2017}, and computed the detection efficiency of our RV-variability selection. For each target, assuming circular orbits, we used the actual observing sequence with the real epochs, exposure windows, and RV uncertainties, and applied the same $\chi^{2}$ test and $-\log p > 3$ threshold as for the real data. The visible DB WD was assigned $M_{1} = 0.6\,M_{\odot}$, close to the mean mass of DB WDs, and we considered companion
masses of $M_{2} = 0.3$, 0.6, and $1.0\,M_{\odot}$. The semi-amplitude of the
visible star follows from Kepler's third law,
\begin{equation}
	K_{1} = \left(\frac{2\pi G}{P}\right)^{1/3}
	\frac{M_{2}\sin i}{(M_{1}+M_{2})^{2/3}}.
	\label{eq:k1}
\end{equation}
For example, adopting $M_{1}=M_{2}=0.6\,M_{\odot}$ and $i=90^{\circ}$, this yields
$K_{1}\approx330$, $150$, and $110~\mathrm{km\,s^{-1}}$ for
$P=1$~h, $10$~h, and $1$~d, respectively. Inclinations were drawn
from an isotropic distribution and orbital phases at random, with 256
realisations per target and period. 
%The finite exposure time was taken into
%account by averaging the orbital velocity over each exposure, and spectra
%combined from several sub-exposures were modelled as the average of their individual exposures. All times were converted to BJD(TDB). This simulation
%requires the timing of every individual exposure, and is therefore restricted
%to the 444 targets (1666 epochs) observed by DESI and SDSS-V; the legacy SDSS
%and LAMOST spectra are co-adds whose individual exposures, often spread over
%several hours or nights, cannot be reconstructed. Because these co-adds mainly
%extend the long time baselines, the period range of maximum sensitivity derived
%here is not expected to change. Simulations without injected orbits yield a
%false-positive rate of 0.09\%, consistent with our threshold.

\begin{figure*}[ht!]
	\centering
	\includegraphics[width=\textwidth]{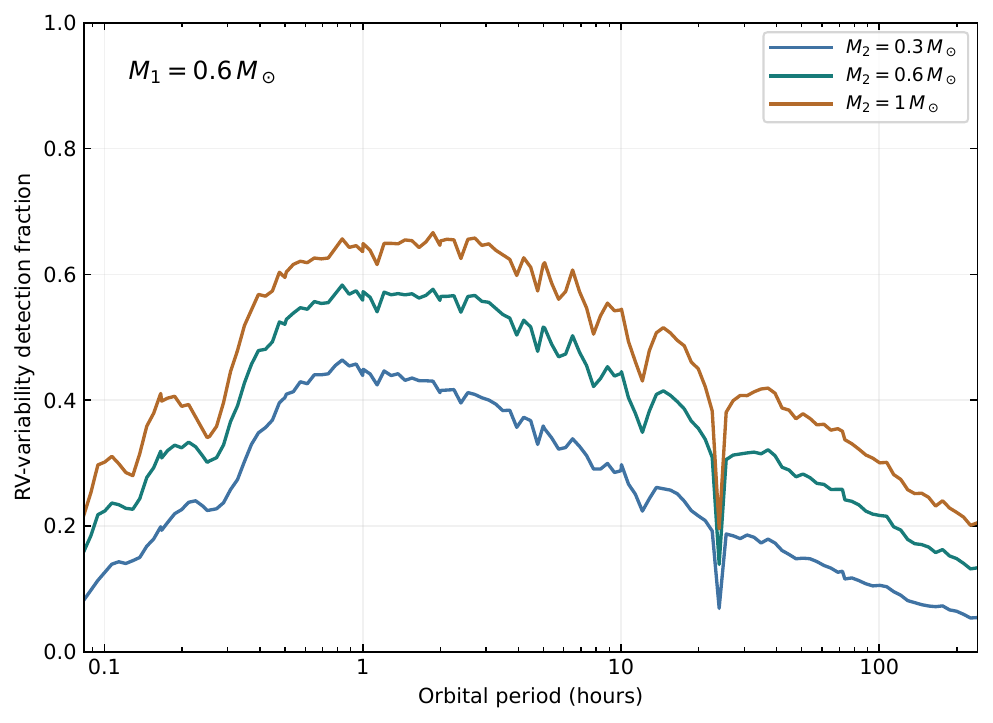}
	\caption{Fraction of simulated binaries detected as RV variable
		($-\log p > 3$) as a function of orbital period, for a visible DB WD of
		$M_{1} = 0.6\,M_{\odot}$ and companion masses of $M_{2} = 0.3$, 0.6,
		and $1.0\,M_{\odot}$. }
	\label{fig5}
\end{figure*}

The resulting detection efficiencies are shown in Figure~\ref{fig5}. For
$M_{1} = M_{2} = 0.6\,M_{\odot}$, the efficiency is highest
($\simeq$50--58\%) for $0.5 \lesssim P \lesssim 7$~h, and stays above
$\sim$40\% up to $P \simeq 15$~h. Towards shorter periods it decreases because
the orbital motion is averaged within a single exposure, reaching
$\simeq$30\% at $P = 15$~min, while towards longer periods it decreases with the
velocity amplitude, to $\simeq$26--31\% at 1--3~d and $\simeq$13\% at 10~d. The
narrow dip at $P \simeq 24$~h is caused by epochs separated by nearly integer
numbers of days, which sample the same orbital phase. A more massive companion
raises the efficiency and broadens the sensitive range (above 50\% for
$0.35 \lesssim P \lesssim 15$~h when $M_{2} = 1.0\,M_{\odot}$), whereas for
$M_{2} = 0.3\,M_{\odot}$ it does not exceed $\sim$46\%. The efficiency remains
well below unity because about 40\% of our targets have only two epochs and
because of unfavourable inclinations and orbital phases. Our search is therefore most sensitive to DB
binaries with orbital periods of $\sim$0.5--10~h. 

\section{Conclusion}
\label{sec:conclusion}

We have presented the first systematic search for RV-variable DB WDs and
DB DWD binary candidates. Our initial sample was taken from the DB WD
catalogue classified from DESI DR1 by \citet{swan2026desidatarelease1}, and was
crossmatched with LAMOST DR13 and SDSS DR20 within 3~arcsec to obtain
multi-epoch spectroscopy.

We measured RVs from the LAMOST and SDSS spectra by template fitting,
and adopted the per-exposure RVs of \citet{swan2026desidatarelease1} for DESI after
applying quality cuts. The reliability of the RV uncertainties was
verified using the RV differences of exposure pairs of the same targets.
We then assessed the significance of RV variability with a
$\chi^{2}$-based test, adopting $-\log p > 3$ as the selection criterion.
Among the 906 DB WDs with at least two valid RV measurements, we
identified 9 DB binary candidates, two of which are regarded as
low-confidence candidates because their spectra are contaminated by light
from a likely wide M dwarf companion. A literature check through SIMBAD
shows that none of them has been reported as a binary or binary candidate
before, so all 9 candidates are newly identified.

Monte Carlo simulations based on the actual observing epochs show that our
selection is most sensitive to binaries with orbital periods of
$\sim$0.5--10~h, with detection efficiencies of $\simeq$50--58\% for
$M_{1} = M_{2} = 0.6\,M_{\odot}$. Follow-up spectroscopic observations of these
candidates, ideally consisting of several exposures within a single night, will be
essential to confirm their binary nature and to
determine their orbital periods, which will further constrain the
properties of the DB DWD population and its connection to potential
SN~Ia progenitors and gravitational-wave sources.
%\section{Using Chinese, Japanese, and Korean characters}
%
%Authors have the option to include names in Chinese, Japanese, or Korean (CJK)
%characters in addition to the English name. The names will be displayed
%in parentheses after the English name. The way to do this in AASTeX is to
%use the CJK package available at \url{https://ctan.org/pkg/cjk?lang=en}.
%Further details on how to implement this and solutions for common problems,
%please go to \url{https://journals.aas.org/nonroman/}.

%% For this sample we use BibTeX plus aasjournalv7.bst to generate the

\begin{acknowledgments}
K.Y.L., H.H.Y., J.K.Z and G.Z. acknowledge support from the Strategic Priority Research Program of the Chinese Academy of Sciences (Grant No. XDB1160301) and the National Natural Science Foundation of China (NSFC; Grant No. 12588202 and Grant No. 12273055). H.H.Y. also acknowledges support from the NSFC through Grant No. 12233013 and thanks the Young Elite Scientists Sponsorship Program by the China Association for Science and Technology (CAST) for its support.
\end{acknowledgments}

\bibliography{sample701}{}
\bibliographystyle{aasjournalv7}



\end{document}